\documentclass{aa}
\usepackage{natbib}
\bibpunct{(}{)}{,}{a}{}{,} 
\usepackage{graphicx}

\def\la{\;
\raise0.3ex\hbox{$<$\kern-0.75em\raise-1.1ex\hbox{$\sim$}}\; }
\def\ga{\;
\raise0.3ex\hbox{$>$\kern-0.75em\raise-1.1ex\hbox{$\sim$}}\; }

\newcommand{\dmm}{$\Delta\mu/\mu\,$}
\newcommand{\daa}{$\Delta\alpha/\alpha\,$}
\newcommand{\dff}{$\Delta F/F\,$}
\newcommand{\kms}{km~s$^{-1}\,$}
\newcommand{\ms}{m~s$^{-1}\,$}

\newcommand{\cmm}{cm$^{-3}\,$}
\newcommand{\etal}{{et al.}}

\begin{document}

\title{
Searching for spatial variations of $\alpha^2/\mu$ in the Milky Way
}
\subtitle{}
\author{S. A. Levshakov\inst{1,2,3}
\and
P. Molaro\inst{1}
\and
D. Reimers\inst{4}
}
\institute{
INAF-Osservatorio Astronomico di Trieste, Via G. B. Tiepolo 11,
34131 Trieste, Italy\\
\email{lev@astro.ioffe.rssi.ru} 
\and
Key Lab. for Research in Galaxies and Cosmology,
Shanghai Astronomical Observatory, CAS, 80 Nandan Road, Shanghai 200030,
P.R. China
\and
Ioffe Physical-Technical Institute, 
Polytekhnicheskaya Str. 26, 194021 St.~Petersburg, Russia
\and
Hamburger Sternwarte, Universit\"at Hamburg,
Gojenbergsweg 112, D-21029 Hamburg, Germany
}
\date{Received 00  ; Accepted 00}
\abstract
{}
{
We probe the dependence of $\alpha^2/\mu$
on the ambient matter density by means of submm- and mm-wave bands spectral observations
in the Milky Way.
} 
{A procedure is suggested to explore 
the value of $F = \alpha^2/\mu$, where 
$\mu = m_{\rm e}/m_{\rm p}$ is the electron-to-proton mass ratio, 
and $\alpha = e^2/(\hbar c)$ is the fine-structure constant. 
The fundamental physical constants, which are measured in different physical environments of
high (terrestrial) and low (interstellar) densities of baryonic matter are
supposed to vary in chameleon-like scalar field models, which predict that
both masses and coupling constant may depend on the local matter density.
The parameter $\Delta F/F = (F_{\rm obs} - F_{\rm lab})/F_{\rm lab}$
can be estimated from the radial velocity offset, 
$\Delta V = V_{\rm rot} - V_{\rm fs}$, between
the low-laying rotational transitions in carbon monoxide $^{13}$CO and
the fine-structure transitions in atomic carbon [\ion{C}{i}]. 
A model-dependent constraint on \daa\ can be obtained from \dff\
using \dmm\ independently measured from the ammonia method.
}
{
Currently available radio astronomical datasets provide an upper limit on 
$|\Delta V| < 110$ \ms\ ($1\sigma$). 
When interpreted in terms of the spatial variation of $F$, this gives
$|\Delta F/F| < 3.7\times10^{-7}$. 
An order of magnitude improvement of this limit will allow us to test
independently a non-zero value of 
\dmm = $(2.2\pm0.4_{\rm stat}\pm0.3_{\rm sys})\times10^{-8}$ 
recently found with the ammonia method.
Taking into account that the ammonia method restricts the spatial variation of $\mu$
at the level of $|\Delta\mu/\mu| \leq 3\times10^{-8}$ 
and assuming that \dff\ is the same in the entire interstellar medium,
one obtains that the spatial variation of $\alpha$ does not exceed the value
$|\Delta\alpha/\alpha| < 2\times10^{-7}$. 
Since extragalactic gas clouds have densities similar
to those in the interstellar medium, the bound on \daa\ 
is also expected to be less than $2\times10^{-7}$ at high redshift 
if no significant temporal dependence of $\alpha$ is present.
}
{}
\authorrunning{S. A. Levshakov \etal}
\titlerunning{Searching for spatial variations of $\alpha^2/\mu$ in the Milky Way}

\maketitle

\section{Introduction}
\label{sect-1}

The dimensionless physical constants like
the electron-to-proton mass ratio, $\mu = m_{\rm e}/m_{\rm p}$,
or the fine-structure constant, $\alpha = e^2/(\hbar c)$,
are expected to be dynamical quantities in modern extensions of 
the standard model of particle physics  
(Uzan 2003;  Garcia-Berro \etal\ 2007; Martins 2008; Kanekar 2008; Chin \etal\ 2009).
Exploring these predictions is a subject of many high precision measurements
in contemporary laboratory and astrophysical experiments.
The most accurate laboratory constraints on temporal $\alpha$- and $\mu$-variations 
of
$\dot{\alpha}/\alpha = (-1.6\pm2.3)\times10^{-17}$ yr$^{-1}$,
and
$\dot{\mu}/\mu = (1.6\pm1.7)\times10^{-15}$ yr$^{-1}$ 
were obtained by Rosenband \etal\ (2008), and
Blatt \etal\ (2008), respectively.

In case of monotonic dependence of $\alpha(t)$ and $\mu(t)$ on cosmic time, 
at redshift $z \sim 2$ (corresponding look-back time is $\Delta t \sim 10^{10}$ yr)
the changes of $\alpha$ and $\mu$ would be restricted at the level of
$|\Delta\alpha/\alpha| < 4\times10^{-7}$ and
$|\Delta\mu/\mu| < 3\times10^{-5}$. 
Here \daa\ (or \dmm) is a fractional change in $\alpha$ between a reference value $\alpha_1$ and
a given measurement $\alpha_2$ obtained at different epochs or at different spatial coordinates:
$\Delta\alpha/\alpha = (\alpha_2 - \alpha_1)/\alpha_1$.  

These constraints are in line with geological measurements of
relative isotopic abundances in the Oklo natural fission reactor 
which allows us to probe $\alpha(t)$ at $\Delta t \sim 2\times10^9$ yr ($z \sim 0.4$).
Assuming possible changes only in the electromagnetic coupling constant, Gould \etal\ (2006)
obtained a model dependent constraint on $|\Delta\alpha/\alpha| < 2\times10^{-8}$.
However, when the strength of the strong interaction,~-- the parameter $\Lambda_{QCD}$,~--
is also suggested to be variable, the Oklo data does not provide any bound on
the variation of $\alpha$ (Flambaum \& Shuryak 2002; Chin \etal\ 2009).  

Current astrophysical measurements at higher redshifts are as follows.
There was a claim for a variability of $\alpha$ at
the 5$\sigma$ confidence level:
$\Delta\alpha/\alpha = -5.7\pm1.1$ ppm 
(Murphy \etal\ 2004)\footnote{Hereafter, 1 ppm = $10^{-6}$.},
but this was not confirmed in other measurements which led to the upper bound
$|\Delta\alpha/\alpha| < 2$ ppm (Quast \etal\ 2004; 
Levshakov \etal\ 2005; Srianand \etal\ 2008; Molaro \etal\ 2008a). 

Measurements of the cosmological $\mu$-variation exhibit a similar tendency. 
Non-zero values of $\Delta\mu/\mu = -30.5\pm7.5$ ppm, 
$\Delta\mu/\mu =  -16.5\pm7.4$ ppm 
(Ivanchik \etal\ 2005), and $\Delta\mu/\mu = -24\pm6$ ppm 
(Reinhold \etal\ 2006) found at $z = 2.595$ (Q 0405--443) and $z = 3.025$ (Q 0347--383) 
from the Werner and Lyman bands of H$_2$
were later refuted by Wendt \& Reimers (2008), King \etal\ (2008)
and Thompson \etal\ (2009) who used the same optical absorption-line spectra of quasars
and restricted changes in $\mu$ at the level of $|\Delta\mu/\mu| < 6$ ppm.
The third H$_2$ system at $z = 2.059$ towards the quasar J2123--0050 also
does not show any evidence for cosmological variation in $\mu$:
$\Delta\mu/\mu = -5.6\pm5.5_{\rm stat}\pm2.9_{\rm sys}$ ppm (Malec \etal\ 2010).  
More stringent constraints were obtained at lower redshifts 
from radio observations of the absorption lines of NH$_3$ and other molecules:
$|\Delta \mu/\mu| < 1.8$ ppm at $z = 0.68$ (Murphy \etal\ 2008), and
$|\Delta \mu/\mu| < 0.6$ ppm at $z = 0.89$ (Henkel \etal\ 2009).
Two cool gas absorbers at $z = 1.36$ (Q 2337--011) and $z = 1.56$ (Q 0458--020)
were recently studied in the \ion{H}{i} 21cm and \ion{C}{i}$\lambda\lambda1560, 1657$
absorption lines providing a constraint on the variation of the product  
$X = g_{\rm p}\alpha^2\mu$ (here $g_{\rm p}$ is the proton gyromagnetic ratio):
$\Delta X/X = -6.8\pm1.0_{\rm stat}\pm6.7_{\rm sys}$ ppm (Kanekar \etal\ 2010).  
Thus, the most accurate astronomical estimates restrict cosmological  
variations of the fundamental physical constants at the level of $\sim$1-2 ppm. 

The estimate of fractional changes in \daa\ and \dmm\ by 
spectral methods is always a measurement of the relative Doppler shifts between
the line centers of different atoms/molecules 
and their comparison with corresponding laboratory values 
(Savedoff 1956; Bahcall \etal 1967; Wolfe \etal\ 1976; Dzuba 1999, 2002; Levshakov 2004;
Kanekar \& Chengalur 2004).
To distinguish the line shifts due to radial motion of the object from those caused by
the variability of constants, lines with different
sensitivity coefficients, ${\cal Q}$, to the variations of $\mu$ and/or $\alpha$ 
are to be used\footnote{${\cal Q}$ is a dimensionless coefficient showing a
relative change of the atomic transition frequency $\omega_i$ in response to 
a change of the physical constant $F$: $\Delta \omega_i/\omega_i = {\cal Q}_i \Delta F/F$.}.   
It is clear that the larger the difference $|\Delta {\cal Q}|$
between two transitions, the higher the accuracy of such estimates.  

Optical and UV transitions in atoms, ions and
molecular hydrogen H$_2$ have similar
sensitivity coefficients with $|\Delta Q|$ not exceeding 0.05
(Varshalovich \& Levshakov 1993; Dzuba 1999, 2002; Porsev \etal\ 2007).
For atomic spectra, the estimate of \daa\ is given 
in linear approximation ($|\Delta\alpha/\alpha| \ll 1$)  
by (e.g., Levshakov \etal\ 2006):
\begin{equation}
\frac{\Delta \alpha}{\alpha} \approx \frac{(V_2 - V_1)}{2c({\cal Q}_1 - {\cal Q}_2)}
\equiv \frac{\Delta V}{2c\Delta {\cal Q}}\, ,
\label{eq1}
\end{equation}
where $V_1, V_2$ are the radial velocities of two atomic lines, and $c$ is
the speed of light.
It was shown in Molaro \etal\ (2008b) that
the limiting accuracy of the wavelength scale calibration 
for the VLT/UVES quasar spectra at any point within the whole optical domain is about 30 \ms,
which corresponds to the limiting relative accuracy between two
lines measured in different parts of the same spectrum
of about 50 \ms.   
Taking into account that $|\Delta Q| \simeq 0.05$, it follows from Eq.(\ref{eq1})
that the limiting accuracy of $\Delta\alpha/\alpha$ is 2 ppm,  
which is the utmost value that can 
be achieved in observations of extragalactic objects with present optical facilities.

A considerably higher sensitivity to the variation of physical constants
is observed in radio range.
For example, van Veldhoven \etal\ (2004) first showed that 
the inversion frequency of the $(J,K) = (1,1)$ level of
the ammonia isotopologue $^{15}$ND$_3$ has the sensitivity coefficient
${\cal Q}_\mu = 5.6$. 
Compared to optical and UV transitions,
the ammonia method proposed by Flambaum \& Kozlov (2007) 
provides 35 times more sensitive estimate of \dmm\ 
from measurements of the radial velocity offset between
the NH$_3$ $(J,K) = (1,1)$ inversion transition at 23.7 GHz and
low-lying rotational transitions of other molecules co-spatially distributed with
NH$_3$:
\begin{equation}
\frac{\Delta \mu}{\mu} \approx 0.289\frac{\Delta V}{c}\ .
\label{eq2}
\end{equation}
The ammonia method was recently used to explore 
possible spatial variations\footnote{Hereafter, the term `spatial variation'
means a possible change in $\mu$ between its terrestrial and interstellar values.} 
of physical constants
from observations of prestellar molecular cores in the 
Taurus giant molecular cloud (Levshakov \etal\ 2010, hereafter L10), 
the Perseus cloud, the Pipe Nebula, and Infrared dark clouds (Levshakov \etal\ 2008b; 
Molaro \etal\ 2009). 
In contrast to the mentioned above laboratory constraints on {\it temporal} variations, 
this method reveals a tentative {\it spatial} variation of \dmm\ at the level of 
\dmm = $(2.2\pm0.4_{\rm stat}\pm0.3_{\rm sys})\times10^{-8}$ (L10).
The corresponding conservative upper limit in this case is equal to
$|\Delta \mu/\mu| \leq 3\times10^{-8}$.

In the present paper
we consider fractional changes of a combination of two constants
$\alpha^2$ and $\mu$, $F = \alpha^2/\mu$, 
which are estimated from the comparison of transition frequencies
measured in different physical environments of
{\it high} (terrestrial) and {\it low} (interstellar) densities of baryonic matter.
The idea behind this experiment is that some class of scalar field models~---
so-called chameleon-like fields~--- predict the dependence
of both masses and coupling constant on the local matter density
(Olive \& Pospelov 2008; Upadhye \etal\ 2010).
Chameleon-like scalar fields were introduced by
Khoury \& Weltman (2004a,b) and by Brax \etal\ (2004) 
to explain negative results on laboratory searches for the fifth force 
which should arise inevitably from couplings between scalar fields and 
standard model particles. 
The chameleon models assume that a light scalar field acquires both an effective potential
and effective mass because of its coupling to matter that depends on the
ambient matter density. In this way, the chameleon
scalar field may evade local tests of the equivalence principle and fifth force experiments
since the range of the scalar-mediated fifth force for the terrestrial
matter densities is too small to be detected.
Similarly, laboratory tests with atomic clocks for $\alpha$-variations 
are performed under conditions of constant local density and, hence,
they are not sensitive to the presence of the chameleon scalar field (Upadhye \etal\ 2010).
This is not the case for space-based tests, where the matter density is considerably lower,
an effective mass of the scalar field is negligible,
and an effective range for the scalar-mediated force is large.
Light scalar fields are usually attributed to a negative pressure substance
permeating the entire visible Universe and known as dark energy (Caldwell \etal\ 1998).
This substance is thought to be responsible
for a cosmic acceleration at low redshifts, $z \la 1$ (Peebles \& Rata 2003; Brax 2009).

\section{[\ion{C}{i}] and CO lines as probes of $\alpha^2/\mu$}
\label{sect-2}

The variations of the physical constants can be probed through atomic fine-structure
(FS) and molecular rotational transitions (Levshakov \etal\ 2008a; Kozlov \etal\ 2008).
The corresponding lines are observed in submm- and mm-wavelength ranges.
Along with a gain in sensitivity, the use
of such transitions allows us to estimate constants at 
very high redshifts ($z > 5$) which are inaccessible to optical observations. 

Let us consider radial velocity offsets between molecular
rotational and atomic FS lines,
$\Delta V = V_{\rm rot} - V_{\rm fs}$.
The offset $\Delta V$ is related to the parameter \dff\
as follows (Levshakov \etal\ 2008a):
\begin{equation}
\Delta F/F \approx 2\Delta \alpha/\alpha - \Delta \mu/\mu \equiv \Delta V/c\, .
\label{eq3}
\end{equation}
The velocity offset in Eq.(\ref{eq3}) can be represented by the sum of two components
\begin{equation}
\Delta V = \Delta V_f + \Delta V_n,
\label{eq4}
\end{equation}
where $\Delta V_f$ is the shift due to $F$-variation, and $\Delta V_n$
is the Doppler noise~--- a random component caused by possible local offsets
since transitions from different species may arise from slightly different
parts of a gas cloud, at different radial velocities.

The Doppler noise yields offsets which can either mimic or obliterate a real signal.
Nevertheless, if these offsets are of random nature, the 
signal $\Delta V_f$ can be estimated statistically by
averaging over a large data sample:
\begin{equation}
\langle \Delta V \rangle = \langle \Delta V_f \rangle,\\
Var(\Delta V) = Var(\Delta V_f) + Var(\Delta V_n)\, .
\label{eq5}
\end{equation}
Here we assume that the noise component has zero mean and a finite variance.

The Doppler noise component can be minimized if the chosen species are
closely trace each other. An appropriate pair in our case is the
atomic carbon FS transitions and rotational transitions of carbon monoxide $^{13}$CO.
The spatial distributions of $^{13}$CO and
[{C}\,{\sc i}] are known to be well correlated 
(Keene \etal\ 1985; Meixner \& Tielens 1995; Spaans \& van Dishoeck 1997;
Ikeda \etal\ 2002;  Papadopoulos \etal\ 2004).
The carbon-bearing species C$^0$, C$^+$, and CO are observed in photodissociation
regions (PDRs)~-- neutral regions where chemistry and heating are regulated
by the far-UV photons (Hollenbach \& Tielens 1999).
The PDR is either the interface between the \ion{H}{ii} region and the molecular
cloud or a neutral component of the diffuse interstellar medium (ISM). 
Far-UV photons ($6.0$ eV $< h\nu < 13.6$ eV) are produced by OB stars.
Photons with energy greater than 11.1 eV dissociate CO into atomic carbon and oxygen.
Since the C$^0$ ionization potential of 11.3 eV is quite close to the CO
dissociation energy, neutral carbon can be quickly ionized. This suggests the chemical
stratification of the PDR in the line C$^+$/C$^0$/CO with increasing depth from
the surface of the PDR. Then, one can assume that in the outer envelopes
of molecular clouds neutral carbon lies within a thin layer determined by the
equilibrium between photoionization/recombination processes on the C$^+$/C$^0$ side,
and photodissociation/molecule formation processes on the C$^0$/CO side.
However, observations (Keene \etal\ 1985; Zhang \etal\ 2001)
do not support such a steady-state model which predicts that
C$^0$ should arise only near the edges of molecular clouds. 
To explain the observed correlation between the spatial distributions of
C$^0$ and CO, inhomogeneous PDRs with clumping molecular gas were suggested. 
The revealed ubiquity of the [\ion{C}{i}] transition
$^3P_1 \rightarrow$ $^3P_0$ in molecular clouds is in agreement with
clumpy PDR models (Meixner \& Tielens 1995; Spaans \etal\ 1997;
Papadopoulos \etal\ 2004).

The ground state of the C$^0$ atom consists of the $^3P_{1,2,3}$ triplet levels.
The energies of the fine-structure excited levels relative to the ground state
are $E_{0,1} = 24$ K, and $E_{0,2} = 63$ K, and the transition
probabilities are $A_{1,0} = 7.932\times10^{-8}$ s$^{-1}$, and
$A_{2,1} = 2.654\times10^{-7}$ s$^{-1}$ (Silva \& Viegas 2002).
The excitation rates of the [\ion{C}{i}] $J = 1$ and $J = 2$ levels for
collisions with H$_2$ at $T_{\rm kin} \sim 20$ K are
$q_{0,1} \approx q_{0,2} \approx 10^{-10}$ cm$^3$s$^{-1}$ (Schr\"oder \etal\ 1991).
This implies that for the $J = 1$ and $J = 2$ levels the critical densities
are 1000 \cmm\ and 3000 \cmm, respectively. The low-$J$ rotational transitions
of CO trace similar moderately dense ($n \sim 10^3$ \cmm) and cold ($T_{\rm kin} \sim 20$ K)
gas. It is not completely excluded, however, that 
some heterogeneity of spatial distributions of [\ion{C}{i}] and $^{13}$CO may occur
resulting in the radial velocity offsets. 
 
In the chameleon-like scalar field models for density-dependent $\mu(\rho)$ and $\alpha(\rho)$
the fractional changes in these constants arise from the shift in the expectation value
of the scalar field between high and low density environments.
Since the matter density in the interstellar clouds is $\sim 10^{16}$ times
lower than in terrestrial environments, whereas  
gas densities between the molecular clouds themselves are much smaller
($n_{\rm gas} \sim 10^3-10^5$ \cmm), 
all interstellar clouds 
can be considered as having similar physical conditions
irrespective of their location in space. 
This means that the noise component in Eq.(\ref{eq5}) can be
reduced by averaging over individual \dff\ values obtained 
from an ensemble of clouds for which the measurements of both
[\ion{C}{i}] and $^{13}$CO lines are available. 

Equations~(\ref{eq2}) and (\ref{eq3}) show that 
in order to estimate \dff\ and \dmm\ with a comparable relative error
the uncertainty of the velocity offset in (\ref{eq3}) must be
$\sim$3.5 times smaller than that in the ammonia method ($\sim$5 \ms, see L10).
At the moment such data do not exist. Both laboratory and astronomical
measurements of the [{C}\,{\sc i}] frequencies  
have much larger uncertainties.
For example, the rest frequencies of the 
[{C}\,{\sc i}] $J=1-0$ transition 492160.651(55) MHz (Yamamoto \& Saito 1991)
and $J=2-1$ transition 809341.97(5) MHz (Klein \etal\ 1998) 
are measured with the uncertainties of 
$\varepsilon_v = 33.5$ \ms\ and 18.5 \ms, respectively.
For $^{13}$CO 
the rest hyper-fine frequencies of low-$J$ rotational transitions 
are known with good accuracy:
$\nu_{1-0} = 110.201354280(37)$ GHz, and 
$\nu_{2-1} = 220.398684129(66)$ GHz, i.e.,
$\varepsilon_v \la 0.1$ \ms\ (Cazzoli \etal\ 2004).
Suggesting that the laboratory error $\varepsilon_v \simeq 34$ \ms\ dominates over the errors 
from $\Delta V$ measurements, one obtains that the $\Delta F/F$ 
limiting accuracy is $0.1$ ppm.
To put in another words, if both species arise from the same volume
elements and their radial velocities are known with a typical error of
$\sim 100$ \ms\ (e.g., Ikeda \etal\ 2002), then the mean $\Delta V$ can be estimated with a
statistical error of $\sim 30$ \ms\
from an ensemble of $n \sim 20$ independent measurements. 

Unfortunately, 
available observational data do not allow us to probe \dff\ at the 0.1 ppm level.
First at all, only a handful of sources are known where both [{C}\,{\sc i}]
and $^{13}$CO radial velocities were measured 
(Schilke \etal\ 1995; Stark \etal\ 1996; Ikeda \etal\ 2002; Mookerjea \etal\ 2006a,b).
The line profiles from these observations
were usually fitted with single Gaussians
in spite of apparent asymmetries seen in some cases (e.g., Fig.~7 in Ikeda \etal\ 2002).
Besides, the measured radial
velocities were not corrected for different beamsizes. 
As a result, the scatter in $\Delta V$ becomes large, and the accuracy of the
\dff\ estimate deteriorates.

\section{The $\alpha^2/\mu$ estimate}
\label{sect-3}

In this section we consider constraints on the spatial variations of $\alpha^2/\mu$
which can be obtained from observations of emission lines of atomic carbon and
carbon monoxide in submm- and mm-wave regions.
The FS [\ion{C}{i}] lines and low-$J$ rotational lines of $^{13}$CO are observed
towards many galactic and extragalactic objects (Bayet \etal\ 2006; Omont 2007).
For our purpose we selected a few molecular clouds located at different galactocentric
distances where the radial velocities of these species were measured with a sufficiently
high precision ($\varepsilon_v \sim 100$ \ms).

Table~\ref{tbl-1} lists molecular clouds with both [\ion{C}{i}] and $^{13}$CO line
measurements which are available in literature. The data were obtained
under the following conditions.

{\it TMC-1}~--- the Taurus Molecular Cloud ($D \sim 140$ pc). This dark molecular
cloud was studied with the Caltech 10.4m submillimeter telescope on Mauna Kea, Hawaii
(Schilke \etal\ 1995). The beamsize at the [\ion{C}{i}] (1-0) frequency was $15''$,
while at the $^{13}$CO (2-1) frequency it was about $30''$.
Schilke \etal\ observed similar shapes of the [\ion{C}{i}] (1-0) and $^{13}$CO (2-1)
profiles at five positions perpendicular to the molecular ridge close to the
cyanopolyyne peak. The line parameters listed in Table~\ref{tbl-1} were derived by
Gaussian fits, although the line shapes were not exactly Gaussians. 
Therefore the errors of the line parameters are the formal $1\sigma$ errors of the
fitting procedure. 

{\it L183}~--- is an isolated quiescent dark cloud at a distance of about 100 pc
(Mattila 1979; Franco 1989). The observations of the [\ion{C}{i}]  and $^{13}$CO
lines at six positions along an east-west strip through the center of the cloud
were obtained with the 15m James Clerk Maxwell Telescope (JCMT) on Mauna Kea,
Hawaii (Stark \etal\ 1996).
The beamsize at 492 GHz was $10''$ and $22''$ (A-band) and $15''$ (B-band) at 220 GHz. 
The [\ion{C}{i}] and $^{13}$CO data were smoothed to a resolution of 0.4 \kms\ and
0.2 \kms, respectively.
These emission lines show similar asymmetric profiles which can be attributed to two
kinematically different components closely spaced in velocity with central velocities
around 1 \kms\ and 2 \kms. These components are marginally resolved in the [\ion{C}{i}] 
spectra at two positions (\# 7 and 8 in Table~\ref{tbl-1}).
But since $^{13}$CO lines were not resolved at these positions, we include in
Table~\ref{tbl-1} the results of one component Gaussian fits of both 
$^{13}$CO (2-1) and [\ion{C}{i}] (1-0) spectra from Stark \etal\ (1996).

{\it Ceph B}~--- is a giant Cepheus molecular cloud at a distance of $\sim 730$ pc
located to the south of the Cepheus OB3 association of early-type stars (Blaauw 1964).
Cepheus B, the hottest $^{12}$CO component of this complex (Sargent 1977, 1979),
is surrounded by an ionization front driven by the UV radiation from the brightest
members of the OB3 association (Felli \etal\ 1978).
The observations of the [\ion{C}{i}] (1-0) line were obtained using the KOSMA 3m 
submillimeter telescope on Gornergrat, Switzerlaand (Mookerjea \etal\ 2006a).
This dataset was complemented with $^{13}$CO observed with the IRAM 30m telescope
(Ungerechts \etal\ 2000).
All data were smoothed to the spatial resolution of $1'$ and the velocity resolution
of 0.8 \kms. Table~\ref{tbl-1} includes [\ion{C}{i}] and $^{13}$CO (2-1) lines 
arising around $V_{\rm LSR}$ of $-13.8$ \kms\ at the position of the hotspot in Cepheus B.
The $V_{\rm LSR}$ values of the [\ion{C}{i}] (1-0) and $^{13}$CO (2-1) positions 
derived from Gaussian fitting were reported in Table~2 of Mookerjea \etal\ (2006a)
without their errors.
However, since the lines look symmetric (Fig.~3, Mookerjea \etal\ 2006a), 
we assign them an error of 0.1 \kms. This is slightly larger than the
uncertainty of $\sim$1/10$th$ of the resolution element,~-- a typical error
of the line position for symmetric profiles,~-- but does not affect significantly
the sample mean value of $\Delta V$.

{\it Orion A,B}~--- are giant molecular clouds located at $\sim 450$ pc (Genzel \& Stutzki 1989).
The observations of the [\ion{C}{i}] (1-0) line towards 9 deg$^2$ area of the Orion A cloud and
6 deg$^2$ area of the Orion B cloud with a grid spacing of $3'$ were carried out with the 1.2m
Mount Fuji submillimeter telescope (Ikeda \etal\ 2002). 
These observations were complemented with the $^{13}$CO (1-0) dataset 
presented in Table~3 in Ikeda \etal.
At the frequency 492 GHz the spatial and velocity resolutions were,
respectively, $2.2'$ and 1.0 \kms, whereas at the frequency 110 GHz 
they were $1.6'$ and 0.3 \kms.
The profiles of the [\ion{C}{i}] (1-0) and $^{13}$CO (1-0) lines were found to be
very similar. All spectra were well fitted with one or two
Gaussian functions, and the velocity centers of the [\ion{C}{i}] and $^{13}$CO
lines are almost the same: $|\Delta V| = 0.2\pm0.1$ \kms. 
The results of the Gaussian fitting are given in Table~\ref{tbl-1}.

{\it Cas A}~--- is a supernova remnant at a distance of $\sim 3$ kpc
(Braun \etal\ 1987).  It was mapped in the [\ion{C}{i}] (1-0) line on 
the KOSMA 3m submillimeter telescope
with the beamwidth of $55''$ and the velocity resolution of 0.6 \kms\ 
(Mookerjea \etal\ 2006b). These observations have been compared with 
the $^{13}$CO (1-0) observations (beamsize $\sim 60''$, spectral resolution
$\sim 0.1$ \kms) taken from Liszt \& Lucas (1999).
Both the [\ion{C}{i}] (1-0) and $^{13}$CO (1-0) emission spectra were averaged
over the disk of Cassiopeia A. The results of Gaussian fitting of subcomponents
resolved in the [\ion{C}{i}] (1-0) and $^{13}$CO (1-0) spectra are included in
Table~\ref{tbl-1}. Two strong emission feature observed in both [\ion{C}{i}]
and $^{13}$CO (1-0) lines were identified with the Perseus arm at $-47$ \kms\
($\sim 2$ kpc distant) and with the local Orion arm at $-1$ \kms ($\sim 460$ pc distant).

\begin{table*}[t!]
\centering
\caption{Parameters derived from Gaussian fits to the $^{13}$CO $J=2-1, J=1-0$ and
[C\,{\sc i}] $J=1-0$ emission line profiles observed towards Galactic molecular clouds.
$V_{\scriptscriptstyle LSR}$ is the line center,
$\sigma_v$ is the line width (FWHM).
The numbers in parentheses are the standard deviations in units of the last
significant digit (see text for more details).
Col.~7 lists velocity offsets
$\Delta V = V_{\scriptscriptstyle LSR}^{\scriptscriptstyle CO}
-V_{\scriptscriptstyle LSR}^{\scriptscriptstyle C\,I}$,
and their estimated errors.
}
\label{tbl-1}
\begin{tabular}{l c r@{.}l r@{.}l r@{.}l r@{.}l r@{.}l l}
\hline
\hline
\noalign{\smallskip}
\multicolumn{1}{c}{Source} & No. &
\multicolumn{2}{c}{$V_{\scriptscriptstyle LSR}^{\scriptscriptstyle CO}$,} &
\multicolumn{2}{c}{$\sigma^{\scriptscriptstyle CO}_v$,} &
\multicolumn{2}{c}{$V^{\scriptscriptstyle C\,I}_{\scriptscriptstyle LSR}$,} &
\multicolumn{2}{c}{$\sigma^{\scriptscriptstyle C\, I}_v$,} &
\multicolumn{2}{c}{$\Delta V,$} & \multicolumn{1}{c}{Ref.} \\
 & & \multicolumn{2}{c}{\kms} & \multicolumn{2}{c}{\kms} &
\multicolumn{2}{c}{\kms}&\multicolumn{2}{c}{\kms}
&\multicolumn{2}{c}{\kms}&\\[-2pt]
\multicolumn{1}{c}{$\scriptscriptstyle (1)$} &
\multicolumn{1}{c}{$\scriptscriptstyle (2)$} &
\multicolumn{2}{c}{$\scriptscriptstyle (3)$} &
\multicolumn{2}{c}{$\scriptscriptstyle (4)$} &
\multicolumn{2}{c}{$\scriptscriptstyle (5)$} &
\multicolumn{2}{c}{$\scriptscriptstyle (6)$} &
\multicolumn{2}{c}{$\scriptscriptstyle (7)$} &
\multicolumn{1}{c}{$\scriptscriptstyle (8)$} \\
\noalign{\smallskip}
\hline
\noalign{\medskip}
TMC-1&1 & 6&1(1)$^a$&2&0(1)&6&0(1)&1&5(2)&0&1(1)& Schilke \etal\ 1995\\[-2pt]
     &2 & 6&1(1)$^a$ & 1&6(1) & 6&1(1) & 1&5(1) & 0&0(1) & \\[-2pt]
     &3 & 6&1(1)$^a$ & 1&7(1) & 5&9(1) & 1&6(1) & 0&2(1) & \\[-2pt]
     &4 & 6&1(1)$^a$ & 1&6(1) & 5&8(1) & 1&2(2) & 0&3(1) & \\[-2pt]
     &5 & 6&2(1)$^a$ & 1&5(1) & 6&4(1) & 2&0(2) & --0&2(1) & \\
\noalign{\smallskip}
L183 &6  & 2&83(3)$^a$ &2&00(7) &2&2(2)&1&7(3) & 0&63(20) & Stark \etal\ 1996 \\[-2pt]
     &7  & 2&41(4)$^a$ &1&81(9) &1&6(2)&2&2(4) & 0&81(20) & \\[-2pt]
     &8  & 2&16(3)$^a$ &1&84(7) &2&2(1)&1&6(3) & --0&04(10) \\[-2pt]
     &9  & 2&05(4)$^a$ & 2&6(1) &1&6(1)&1&8(3) & 0&45(11) & \\[-2pt]
     &10 & 2&20(4)$^a$ & 2&0(1) &2&4(1)&1&4(3) & --0&20(11) & \\
\noalign{\smallskip}
Ceph B&11&--13&4(1)$^a$ & 2&1      &--14&1(1)    &1&9    &0&7(1) & Mookerjea \etal\ 2006a\\
\noalign{\smallskip}
Ori A,B&12&8&6(1)$^b$&4&0(1)&9&3(1)&4&7(1)&--0&7(1)& Ikeda \etal\ 2002\\[-2pt]
       &13&7&1(1)$^b$&2&4(1)&7&6(1)&3&4(1)&--0&5(1)& \\[-2pt]
       &14&4&8(1)$^b$&2&5(1)&5&2(2)&3&3(4)&--0&4(2)& \\[-2pt]
       &15&10&4(1)$^b$&2&9(1)&10&9(3)&3&3(1)&--0&5(3)& \\[-2pt]
       &16&11&2(1)$^b$&2&6(1)&11&0(1)&3&3(1)&0&2(1)& \\[-2pt]
       &17&9&6(1)$^b$ &2&7(1)&9&9(1)&3&2(1)&--0&3(1)& \\[-2pt]
      &18&9&6(1)$^b$&4&4(1)&9&5(1)&5&0(1)&0&1(1)& \\[-2pt]
      &19&8&1(1)$^b$&3&3(1)&8&3(1)&3&7(1)&--0&2(1)& \\[-2pt]
      &20&5&3(1)$^b$&2&3(1)&5&5(1)&3&0(1)&--0&2(1)& \\[-2pt]
      &21&10&3(2)$^b$&3&4(4)&10&6(1)&4&2(1)&--0&3(2)& \\[-2pt]
      &22&10&6(1)$^b$&1&8(1)&10&4(2)&2&3(4)&0&2(2)& \\[-2pt]
      &23&9&6(1)$^b$&2&7(1)&9&8(1)&3&4(1)&--0&2(1)& \\
\noalign{\smallskip}
Cas A &24&--46&80(5)$^b$ & 2&1(1) &--47&4(1)&2&9(2) &0&60(11) & Mookerjea \etal\ 2006b\\[-2pt]
      &25&--1&10(6)$^b$&1&8(1)&--1&7(1)&3&1(4)&0&60(12)& \\
\noalign{\smallskip}
\hline
\noalign{\smallskip}
\multicolumn{13}{l}{$^a$$J=2-1$.\ $^b$$J=1-0$. } 
\end{tabular}
\end{table*}

The velocity offsets $\Delta V$ between the $^{13}$CO
and [C\,{\sc i}] lines are given in Col.~7 of Table~\ref{tbl-1},
the corresponding linewidths, $\sigma_v$, are shown in Cols.~4 and 6.
When both transitions trace the same material, the lighter element \ion{C}
should always have larger linewidth. If the line broadening is caused
by thermal and turbulent motions, i.e.,
$\sigma^2_v = \sigma^2_{\rm therm} + \sigma^2_{\rm turb}$,
then for two species with masses $m_1 < m_2$ we have
\begin{equation}
\sqrt{m_1/m_2} \leq \sigma_{2,v}/\sigma_{1,v} \leq 1\, .
\label{eq6}
\end{equation}
In practice, this inequality is fulfilled only approximately.
Except for the pure thermal and turbulent broadening
there are many other mechanisms which can give rise to the broadening of
atomic and molecular lines. These are saturation broadening (lines have different
optical depths), the presence of unresolved velocity gradients
(nonthermal distribution is not normal),
the increasing velocity dispersion of the nonthermal component with
increasing map size (the higher angular resolution is realized for
the higher frequency transitions), etc.
Thus, the consistency of the apparent linewidths defined by Eq.(\ref{eq6})
is a necessary condition for two species with different masses
to be co-spatially distributed, but is not a sufficient one.

From Table~\ref{tbl-1} it is seen that the inequality (\ref{eq6})
is fulfilled for all selected pairs $^{13}$CO/[\ion{C}{i}]
within the estimated uncertainties of the linewidths.
Thus, the whole sample of $n = 25$ $\Delta V$ values can be used
to estimate \dff.

The averaging of the velocity offsets over the dataset gives the unweighted
mean
$\Delta V_{\rm uw} \equiv \langle V_{LSR}(^{13}{\rm CO}) - V_{LSR}$(\ion{C}{i})$\rangle$ =
$0.046\pm0.083_{\rm stat}$ \kms.
With weights inversionally proportional to the variances, one derives
$\Delta V_{\rm w} = 0.029\pm0.077$ \kms. 
The median of the sample is $\Delta V_{\rm med} = 0.0$ \kms, and the robust $M$-estimate
(L10) is $\Delta V_{\rm M} = 0.022\pm0.082$ \kms.
The statistical error for the mean velocity offset measurement is larger
than that expected from the published values of the statistical errors from
the one component Gaussian fits: the mean error of the individual $\Delta V$
is 0.13 \kms, and the expected error of the mean $\Delta V$ is $\sim$0.026 \kms. 
A possible reason for such a high Doppler noise has been discussed in Sect.~2.
The systematic error in this case is dominated by the uncertainty of the rest frequency
of the [\ion{C}{i}] (1-0) transition, $\varepsilon_v = 33.5$ \ms. 
Thus, taking the $M$-estimate
as the best measure of the velocity offset, we have
$\Delta V_{\rm M} = 0.022\pm0.082_{\rm stat}\pm0.034_{\rm sys}$ \kms,
and the $1\sigma$ upper limit on $|\Delta V| < 0.11$ \kms. 

This estimate restricts the spatial
variability of $F$ at the level of $|\Delta F/F| < 0.37$ ppm.
Recently we obtained a constraint on the spatial change of the
electron-to-proton mass ratio $|\Delta\mu/\mu| \leq 0.03$ ppm
based on measurements in cold molecular cores in the  Milky Way  (L10).
Combining these two upper limits, the fine-structure constant
can be bound as $|\Delta \alpha/\alpha| < 0.2$ ppm.

\section{Conclusion}
\label{sect-4}

The level of 0.2 ppm  represents a model-dependent 
upper limit on the {\it spatial} variations of $\alpha$.
Under model-dependence we assume here 
that both \dff\ and \dmm\ do not change
significantly from cloud to cloud, since
astrophysical measurements of these parameters are made in
low density regions of the interstellar medium
with $\rho_{\rm cloud} \ll \rho_{\rm  terrestrial}$.

For comparison, the upper limit on the {\it temporal} $\alpha$-variation
obtained from high-redshift quasar absorbers is 
$|\Delta \alpha/\alpha| < 2$ ppm (Sect.~1).
If dependence of constants on the ambient matter density dominates over 
temporal (cosmological), as suggested in chameleon-like scalar field models, then 
one may expect that $|\Delta\alpha/\alpha| < 0.2$ ppm at high redshifts as well,
since quasar absorbers have gas densities similar to those in the interstellar clouds.
Taking into account that the predicted changes in $\alpha$ and $\mu$
are not independent and that $\mu$-variations may exceed variations in $\alpha$
(e.g., Calmet \& Fritzsch 2002; Langacker \etal\ 2002; Dine \etal\ 2003;
Flambaum \etal\ 2004),
even a lower bound of $|\Delta\alpha/\alpha| \leq 0.03$ ppm is conceivable
within the framework of the chameleon models.

We note that if a theoretical prediction $|\Delta\alpha/\alpha| \ll |\Delta\mu/\mu|$
is valid, then $\Delta F/F \approx -\Delta\mu/\mu$, and, hence, the $F$-estimate with a further
order of magnitude improvement in sensitivity will provide an
independent test of the tentative change of $\mu$.

The factors limiting accuracy of the current estimate of \dff\ at $z = 0$
are a relatively low spectral resolution of the available observations in submm- and mm-wave bands, 
a rather large uncertainty of the rest frequencies of the [C\,{\sc i}] FS lines,
and a small number of objects observed in both [C\,{\sc i}] and $^{13}$CO transitions. 

Modern telescopes like the recently launched Herschel Space Observatory
can provide for Galactic objects the spectral resolution as high as 30 \ms\
(e.g., the Heterodyne Instrument for the Far Infrared, HIFI, has resolving power
$R = 10^7$).
This means that the positions of the [\ion{C}{i}] FS lines
can be measured with the uncertainty of $\sim$3 \ms.
In the near future, high precision measurements will be also available with the
Atacama Large Millimeter/submillimeter Array (ALMA), the Stratospheric Observatory For
Infrared Astronomy (SOFIA), the Cornell Caltech Atacama Telescope (CCAT) and others. 
Thus, any further advances in exploring $\Delta F/F$ depend crucially on
new laboratory measurements of the [C\,{\sc i}] FS frequencies.
If these frequencies will be
known with uncertainties of a few \ms, then the parameter $\Delta F/F$ can be probed 
at the level of $10^{-8}$ which would be comparable with the non-zero
signal in the spatial variation of the electron-to-proton
mass ratio $\mu$.

\begin{acknowledgements}
We thank our anonymous referee for valuable comments
on the manuscript.  
The project has been supported in part by
DFG Sonderforschungsbereich SFB 676 Teilprojekt C4,
the RFBR grants 09-02-12223 and 09-02-00352, 
by the Federal Agency for Science and Innovations grant
NSh-3769.2010.2,
and by the Chinese Academy of Sciences visiting professorship
for senior international scientists under grant No. 2009J2-6. 
\end{acknowledgements}

\end{document}